\documentclass[aip, rsi, amsmath,amssymb, reprint]{revtex4-1}

\usepackage{graphicx}
\usepackage{dcolumn}
\usepackage{bm}
\usepackage[utf8]{inputenc}
\usepackage[T1]{fontenc}
\usepackage{mathptmx}
\usepackage[english]{babel}
\usepackage{amsmath}
\usepackage{amsfonts}
\usepackage{amssymb}
\usepackage{graphicx}
\usepackage{hyperref}
\usepackage{physics}
\usepackage{pgf}
\usepackage{braket}
\usepackage{soul, color}

\graphicspath{{./Figures/}}

\DeclareUnicodeCharacter{2212}{$-$}

\let\pgfimageWithoutPath\pgfimage 
\renewcommand{\pgfimage}[2][]{\pgfimageWithoutPath[#1]{Figures/#2}}

\begin{document}

\title{Ultrahigh Vacuum Packaging and Surface Cleaning for Quantum Devices}

\author{M. Mergenthaler}
\affiliation{IBM Quantum, IBM Research Europe - Zurich, S\"aumerstrasse 4, 8803 R\"uschlikon, Switzerland}
\author{S. Paredes}
\affiliation{IBM Quantum, IBM Research Europe - Zurich, S\"aumerstrasse 4, 8803 R\"uschlikon, Switzerland}
\author{P. M\"uller}
\affiliation{IBM Quantum, IBM Research Europe - Zurich, S\"aumerstrasse 4, 8803 R\"uschlikon, Switzerland}
\author{C. M\"uller}
\affiliation{IBM Quantum, IBM Research Europe - Zurich, S\"aumerstrasse 4, 8803 R\"uschlikon, Switzerland}
\author{S. Filipp}
\affiliation{IBM Quantum, IBM Research Europe - Zurich, S\"aumerstrasse 4, 8803 R\"uschlikon, Switzerland}
\author{M. Sandberg}
\affiliation{IBM Quantum, IBM T. J. Watson Research Center, Yorktown Heights, NY 10598, USA}
\author{J. Hertzberg}
\affiliation{IBM Quantum, IBM T. J. Watson Research Center, Yorktown Heights, NY 10598, USA}
\author{V. P. Adiga}
\affiliation{IBM Quantum, IBM T. J. Watson Research Center, Yorktown Heights, NY 10598, USA}
\author{M. Brink}
\affiliation{IBM Quantum, IBM T. J. Watson Research Center, Yorktown Heights, NY 10598, USA}
\author{A. Fuhrer}
\email{afu@zurich.ibm.com}
\affiliation{IBM Quantum, IBM Research Europe - Zurich, S\"aumerstrasse 4, 8803 R\"uschlikon, Switzerland}

\date{\today}

\begin{abstract}
We describe design, implementation and performance of an ultra-high vacuum (UHV) package for superconducting qubit chips or other surface sensitive quantum devices. The UHV loading procedure allows for annealing, ultra-violet light irradiation, ion milling and surface passivation of quantum devices before sealing them into a measurement package. The package retains vacuum during the transfer to cryogenic temperatures by active pumping with a titanium getter layer. We characterize the treatment capabilities of the system and present measurements of flux tunable qubits with an average T$_1=84~\mu$s and T$^{echo}_2=134~\mu$s after vacuum-loading these samples into a bottom loading dilution refrigerator in the UHV-package.

%
%
%
%

\end{abstract}
\maketitle

\section{Introduction}
\begin{figure}[b!]
\includegraphics[width=.9\columnwidth]{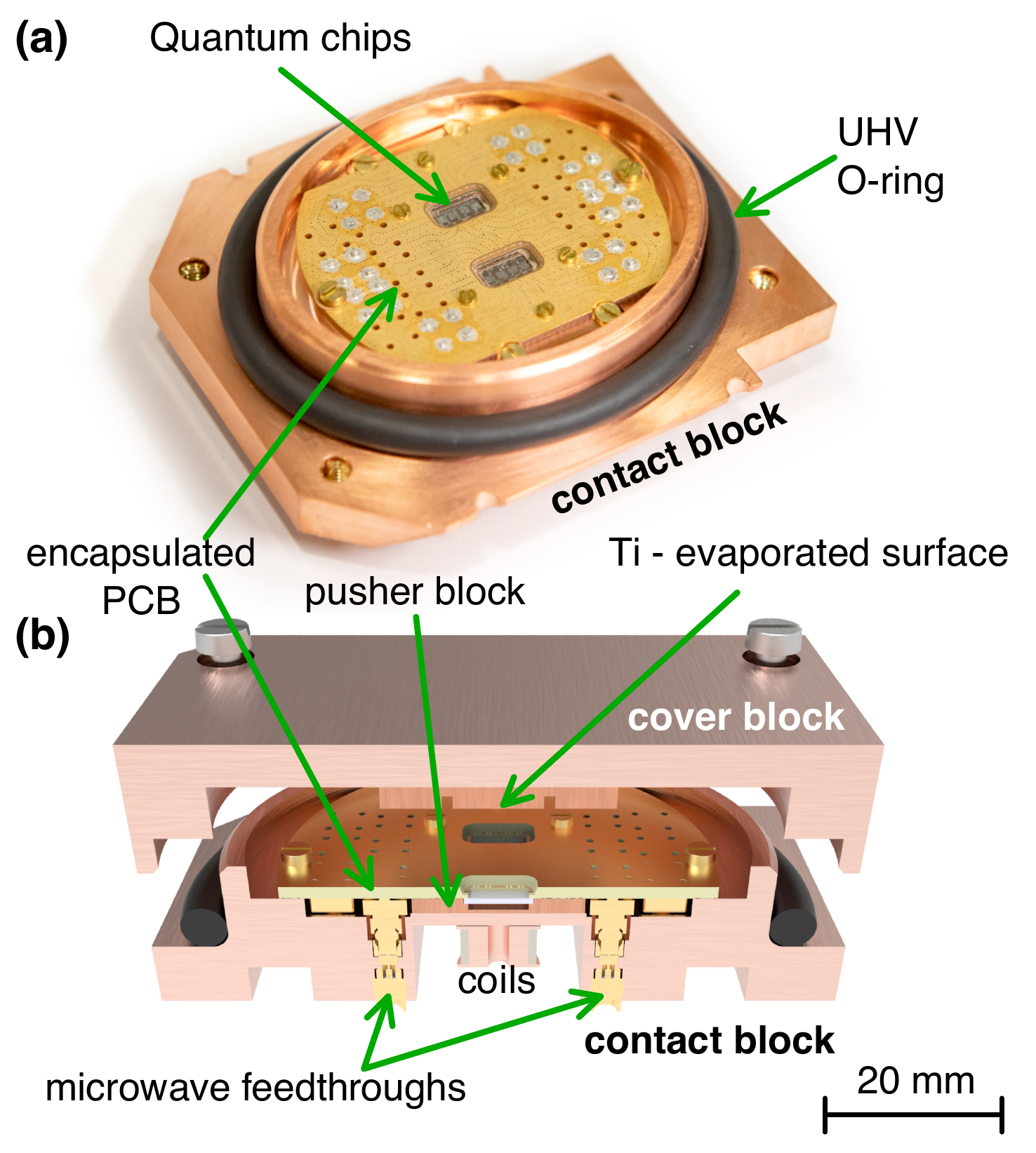}
\caption{\label{fig:fig1} UHV sample package (a) and cross-section (b). The contact block in (a) houses the quantum chips which are clamped below a gold platted PCB. Up to eighteen GPPO vacuum feedthroughs can be used to apply microwave signals. Superconducting coils can be mounted on the outside of the contact block. The cover block makes a vacuum tight seal using a UHV compatible O-ring. }
\end{figure}
The performance of qubits and other quantum components in superconducting quantum processors critically depends on the quality of the used materials~\cite{13OliverAA,17MullerAB,14MartinisAB,13PaladinoAA,05IthierAA}. Disorder, adsorbates, impurities and other unwanted degrees of freedom can lead to qubit relaxation and dephasing by coupling of spurious fluctuators to qubit modes. Material purity and the quality of buried interfaces can be engineered in the fabrication process and typically remains constant after the device is finished. However, the  top surface of planar superconducting qubit chips is much harder to control.  Amorphous oxides, adsorbates or unsaturated surface bonds on the substrate or superconductor surfaces can give rise to fluctuations that manifest themselves as flux and charge noise~\cite{17MullerAB,13PaladinoAA}. These fluctuators are often added through the interaction with the ambient environment after fabrication of the qubit chips and are therefore difficult to avoid or even study.  
In the present paper we demonstrate an ultra-high vacuum (UHV) package that allows selective cleaning or passivation of the top surface of superconducting qubit chips or other surface sensitive quantum devices. Before loading the UHV package, a UHV system facilitates treatment of the sample surface by vacuum outgassing, ultra-violet light irradiation, ion milling or passivation with chemical reagents. The package can then be transferred into a dilution refrigerator and cooled while maintaining vacuum inside the package and thus avoids contamination of the treated surface before measurement. In the following, we first describe the UHV package and UHV treatment system, then we show how the package can be closed in vacuum and how vacuum is maintained during the transfer. The surface treatment capabilities are discussed and finally, measurements on flux-tunable transmon qubits mounted in the UHV package are presented.

\section{Experimental Setup}
\subsection{UHV Package}
The UHV package is shown in Fig.~\ref{fig:fig1}. The bottom contact block has 18 vacuum tight GPPO microwave feedthroughs connected to a PCB that can fit two $4 \times 8$~mm quantum chips. The chips are positioned in a recess of a small copper pusher block, clamped between this block and an eight layer printed circuit board (PCB) and bonded to the PCB using aluminum wire bonds. Other chip sizes can be implemented easily by changing the layout of the PCB. This sandwich of the PCB, chip and pusher block is mounted with brass screws into the contact block and makes contact between the microwave feedthroughs and the PCB mount connectors using bullet adaptors. The shape of the contact block is elongated in order to fit inside the probe of a bottom loading dilution refrigerator (Bluefors LD).  The vacuum seal of the package is made with a UHV compatible O-ring (VAT Valve AG, N-5100-332, 5.33 x 59.69 mm) and was found to be leak tight to a level below $10^{-10}$~mbar~l/sec. This was measured with a standard helium leak tester and a modified cover block that had a ISO-KF 40 flange machined into it. Up to six superconducting coils can be placed below the chips to tune the local flux density e.g. for frequency tunable superconducting qubits. In order to maintain good vacuum during transfer of the package into the measurement apparatus an active getter material such as titanium can be evaporated into the cover block.  This layer then captures molecules outgassing from internal surfaces e.g. from the PCB (see below). After bonding and mounting the two quantum chips, both the contact block and the cover block are loaded into a UHV treatment system where surface treatments can be performed. 

\begin{figure}[t!]
\includegraphics[width=.95\columnwidth]{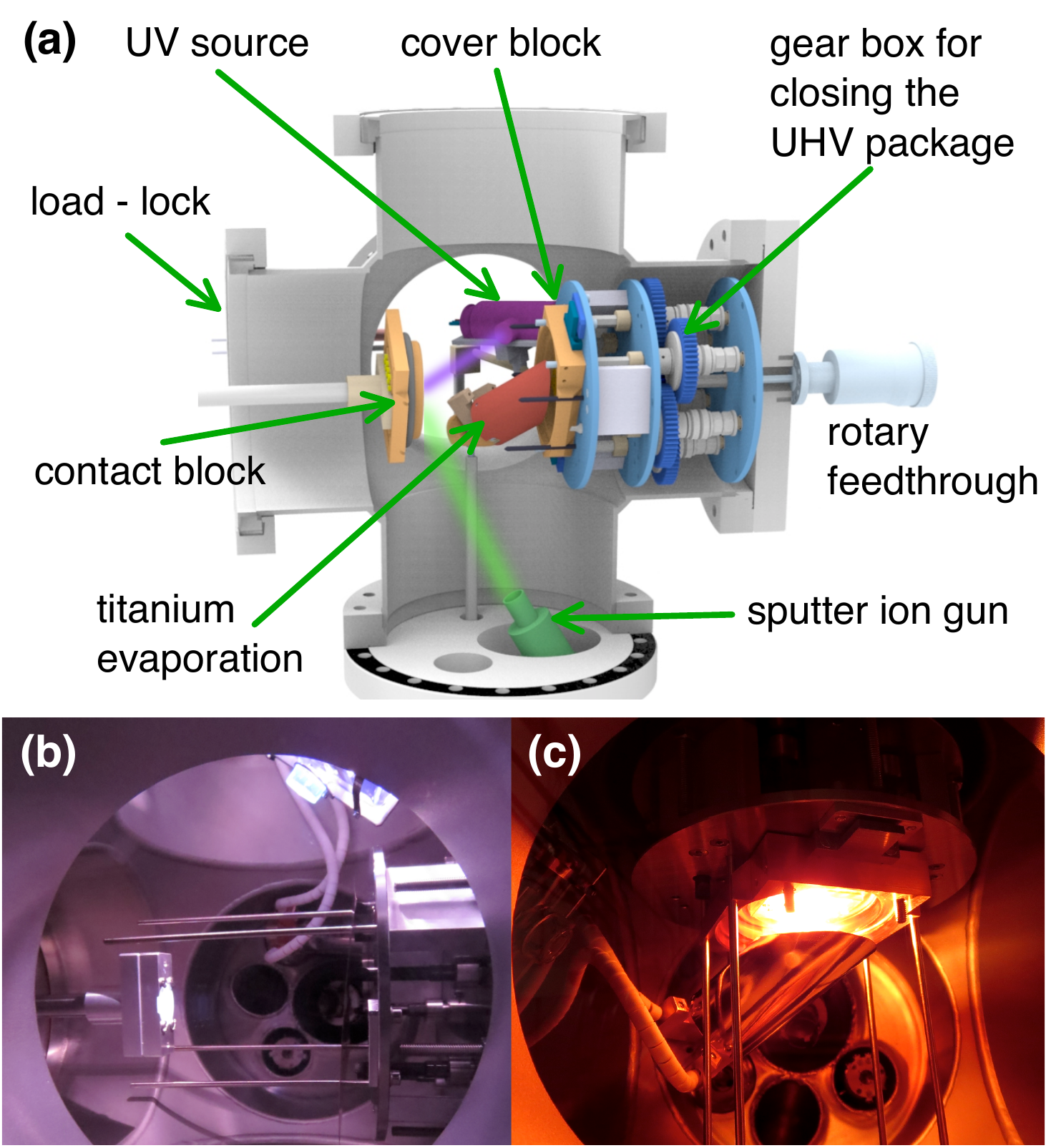}
\caption{\label{fig:fig2} (a) Cross-section of UHV chamber with sputter ion gun, titanium sublimation source and UV light source. The UV source and the sputter ion gun are directed at the sample in the contact block. (b) UV light source (top right) illuminating a phosphorescent circular test sample (lower left). (c) Titanium source during evaporation into the cover block. }
\end{figure}

\subsection{UHV Treatment System}
The UHV treatment system is shown in Fig.~\ref{fig:fig2}(a). It consists of a main treatment chamber and a secondary pumping chamber (not shown). The latter is continuously pumped by an ion getter pump (Perkin Elmer, 400l/s) and a liquid nitrogen cooled titanium sublimation pump (TSP). The UHV package is inserted into the main chamber through a load-lock from the left. Pump-down occurs with a turbo pump (Pfeiffer TMU 261) backed by a dry scroll pump (Edwards XDS10). 

An overnight bake at 70$^{\circ}$C (measured on the contact block) leads to a base pressure of $1.2\times10^{-10}$ mbar when both chambers are linked. Before closing the package with an in-situ mechanical closing mechanism the sample surface can be irradiated with a UV-light source (D$_2$, 30~W, Newport - 63163, see Fig.~\ref{fig:fig2}(b)) or selectively ion milled with an ion bombardment gun (Varian - 981-2043) at an angle of 60$^\circ$ away from the surface normal. The shallow angle of the ion beam and rotation of the contact block allow selective sputtering of adsorbed surface residues or surface oxide films. Fig.~\ref{fig:fig2}(c) shows operation of a custom built titanium sublimation source that is used to deposit titanium into the cover block before closing the sample package. Multiple tantalum shields prevent radiative heating of the surrounding surfaces such that the pressure quickly returns to base pressure. 

\begin{figure}[t!]
\includegraphics[width=.9\columnwidth]{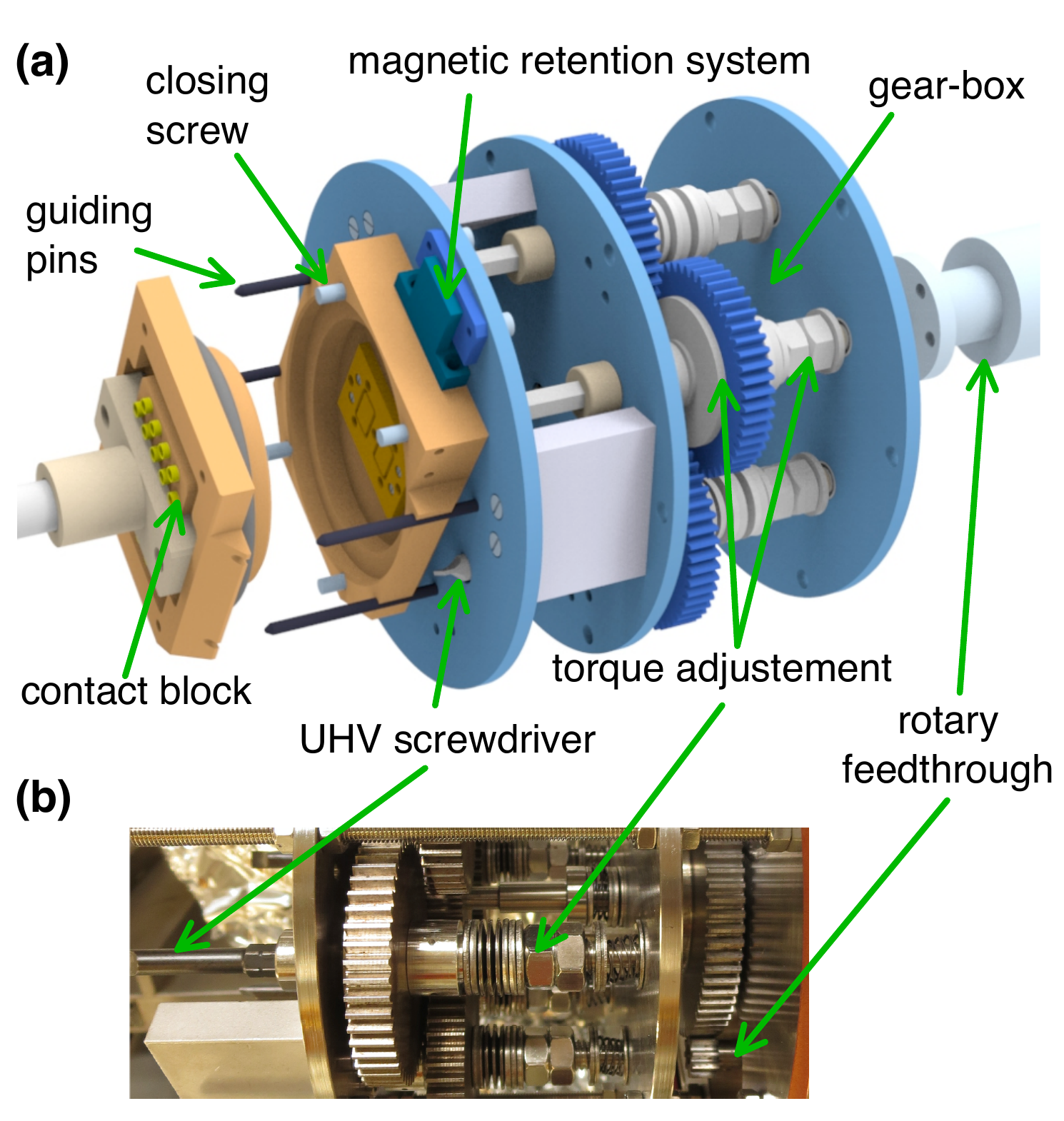}
\caption{\label{fig:fig3} (a) Closing mechanism for in-situ sealing of the UHV package. Each of the six UHV screwdrivers is torque limited and spring loaded such as to allow full closure of the UHV package with four or six screws. The package is held in place by a magnetic retention system. (b) Zoom view of the mechanical vacuum feed-through and torque adjustment of one of the screwdrivers.}
\end{figure}
\subsubsection{Closing of package in UHV}
Figure~\ref{fig:fig3}(a) shows the mechanism that is used for closing the package while maintaining process pressure p~$ \le 1\times10^{-9}$~mbar. The cover block is fixed with a magnetic retention system to this in-situ closing mechanism. Depending on the package, spring-loaded screwdriver heads fit into four or six M4 stainless steel screws. The contact block is aligned with four guiding pins that ensure that the screws fit into the threads on the contact block. The screwdrivers are actuated by a single rotary vacuum feedthrough with a gearbox that allows individual adjustment of the maximum torque for each of the screwdriver heads. The details of the mechanism are shown in Fig.~\ref{fig:fig3}(b). The torque is adjusted by two nuts on each axis which set the pressure on the large gear to the left that receives the rotary motion. If the torque exceeds the limit the corresponding gear turns without rotating the axis with the screwdriver head. This ensures that all screws are tight and the package gets fully closed.   

\begin{figure}[ht!]
\includegraphics[width=.9\columnwidth]{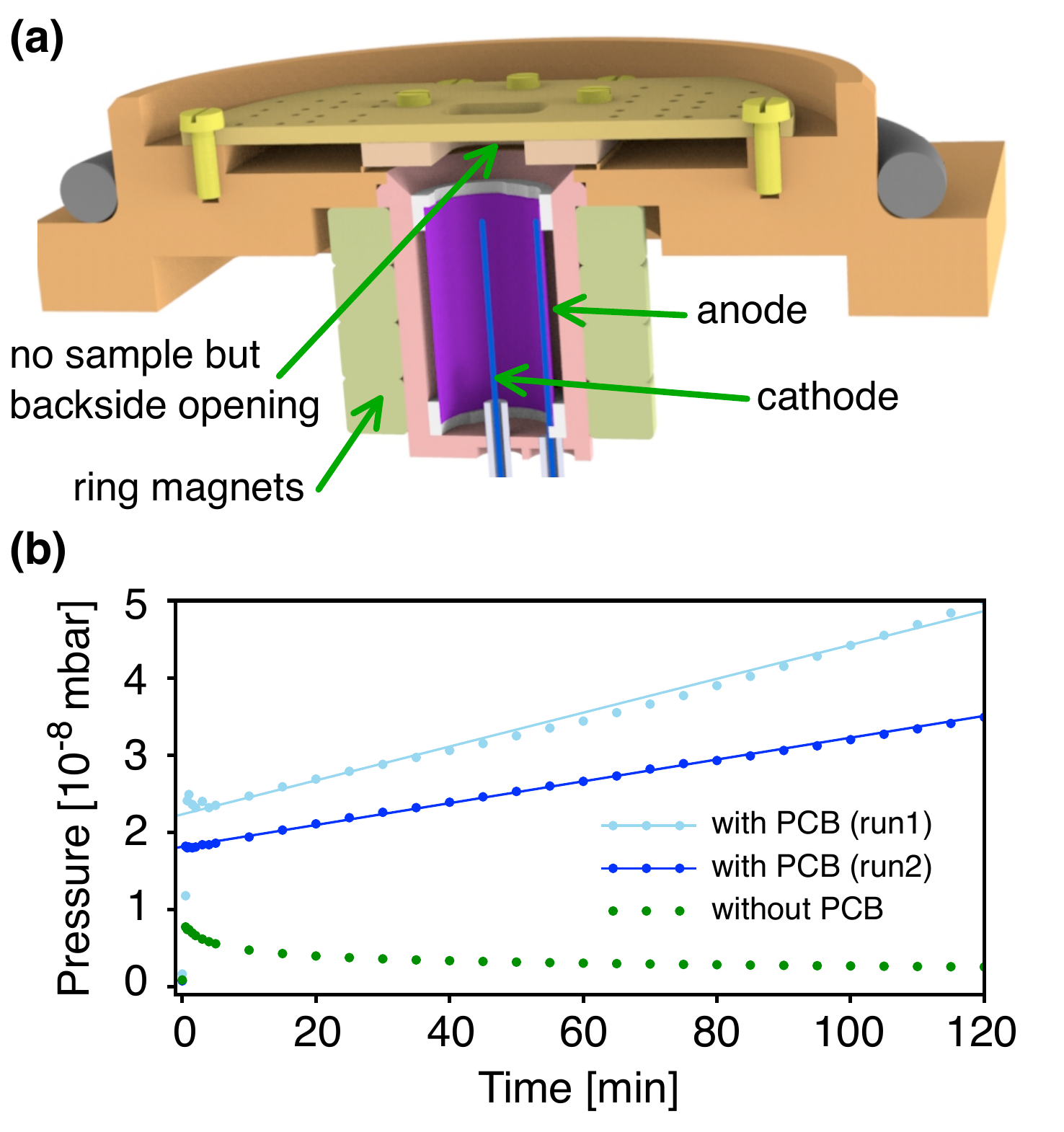}
\caption{\label{fig:fig4} (a) Image of custom built pressure gauge in a typical magnetron geometry. The anode is connected with an isolated tantalum cylinder (purple) and strong ring magnets (green) generate a magnetic field of 0.1~Tesla at the cylinder center.  (b) Pressure change inside the UHV package after titanium evaporation into the cover block and closing of the package. Dots represent measured data points and lines are linear fits for estimates of the leak-rate. }
\end{figure}
 
\subsubsection{UHV conditions during sample transfer}
One of the key advantages of our system is that quantum chips can be loaded into a cryogenic system while maintaining vacuum in the package after surface treatments. It is challenging to fully prove this claim because a pressure measurement near the sample surface is impractical given the small package volume. Instead we implemented a custom built cold cathode magnetron gauge that was mounted on the contact block below the position of the two chips. This is shown as a cross section in Fig.~\ref{fig:fig4}(a). In this test setup there are two holes to the pressure gauge where the samples are normally mounted. The gauge was calibrated with an open package against a commercial cold cathode gauge inside the chamber. It allows us to monitor the pressure inside the package with/without the PCB and with/without titanium getter layer in the cover block for the same duration that is needed to mount and cool the package in our bottom loading dilution refrigerator. A typical transfer of the sample takes about one hour until the package surroundings are pumped down to high vacuum and another hour for the sample to reach a temperature below 70K.

The time dependence of the pressure inside the package is shown in Fig.~\ref{fig:fig4}(b) for two hours after sealing the package. During the sealing process the pressure spikes to a few $1\times10^{-8}$~mbar but without a PCB present in the package the evaporated titanium film again pumps down to a pressure below $3\times10^{-9}$~mbar over the course of an hour (green dots in Fig.~\ref{fig:fig4}(b)). With the PCB mounted in the package we instead find a slight increase of the package pressure with a leak rate of about $2.9\pm0.7\times10^{-14}$~mbar~l/sec given a total inner package volume of 10~ccm. This is shown in Fig.~\ref{fig:fig4}(b) for two separate runs with blue dots. The lines correspond to linear fits with $0.8(1.3)\times10^{-8}$~mbar/h for the dark(light) blue curves respectively. Without titanium evaporation the measured internal leak rate is $3.6\times10^{-11}$~mbar~l/sec (not shown). 

There are a few caveats to these numbers. The internal volume of the contact block in this test setup is larger than for the real package. This may impact the measured internal leak rate negatively due to increased surface desorption and self-outgassing of the gauge. However, our findings indicate that the PCB is the main source of internal outgassing as expected for a standard FR-4 PCB made up of a glass-reinforced epoxy laminate. Furthermore, the pressure is measured on the backside of the PCB where there are additional trapped volumes in our test setup. In the real package the chip surface is only exposed to the titanium film directly above the sample (see Fig.~\ref{fig:fig1}(b)) and we expect the local surface contamination rate to be much smaller than that expected from a pressure of a few $1\times10^{-8}$~mbar. In addition, molecular hydrogen is by far the dominant molecular species in the vacuum system at base pressure. Given the high diffusivity of molecular hydrogen, we expect that it is the main gas species coming out of the PCB laminate and the trapped volumes covered by the PCB. For many surfaces molecular hydrogen is not a contaminant but will nevertheless contribute to the pressure measurement. Finally, we note that outgassing from the PCB could be improved by using a different PCB material such as PTFE or a ceramic.

\subsection{In-situ surface treatments}
 The UHV treatment system offers multiple options to treat the surface of bonded quantum chips before the UHV package is closed.

 \emph{i) Thermal treatment} of the UHV package after loading in order to degas the PCB and other components in the UHV system. We typically perform this step in a 120$^\circ$C bake of the entire vacuum system which heats the loaded chips in the package to about 70$^\circ$C. This allows outgassing of the UV-source, ion bombardment gun and the titanium evaporation source while the entire system is hot and prevents contamination during the subsequent treatment steps. This step also removes most of the physisorbed water and other molecular species on the surfaces of the package and on the chips. It has been shown that thermal treatments at 300$^\circ$C can reduce dielectric loss and fluxnoise in quantum devices~\cite{18GraafAA}. Here we cannot raise the temperature quite as high due to the used materials. This would require a redesign of the PCB and a more local heater.
 
 \emph{ii) Short wavelength UV-irradiation} is known to be efficient in photon induced desorption of various molecular species~\cite{12KoebleyAA,90FoersterAA} including water, H$_2$ and O$_2$. In the treatment system a deuterium lamp with maximum emission between 160 and 300~nm allows irradiation of the chip surface and PCB. A similar treatment was used in a study on SQUIDs and led to a reduction in the measured flux noise~\cite{16KumarAA}. Fig.~\ref{fig:fig2}(c) shows the UHV chamber with a phosphorescent test sample that is illuminated by the UV source. When the UV light is turned on, a pressure spike is typically observed due to the desorption of water and other molecules from the walls of the UHV system. When switching the UV light off after 10~min, pressure quickly relaxes to a level better than before the UV treatment.

\begin{figure}[b!]
\includegraphics[width=.9\columnwidth]{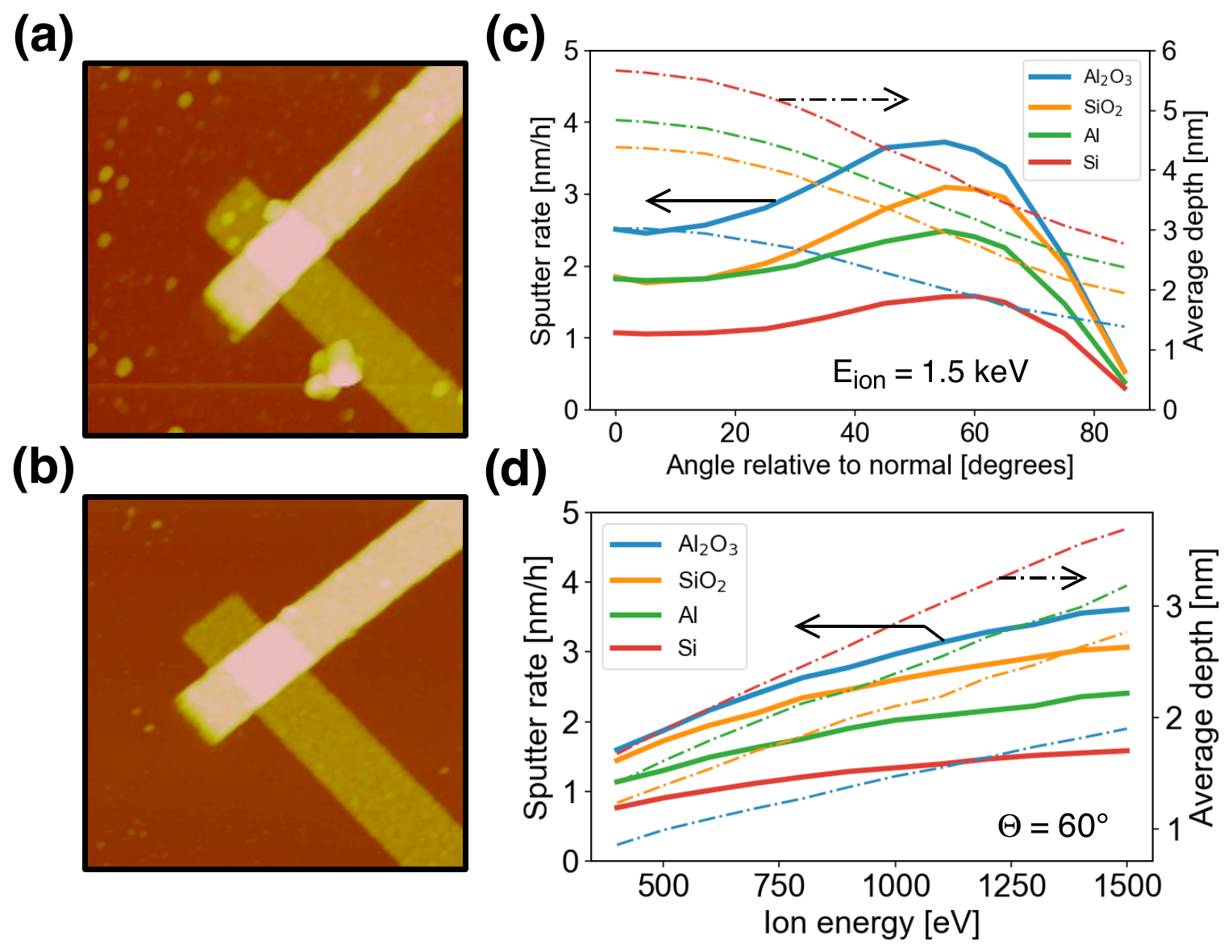}
\caption{\label{fig:fig5} Effectiveness of neon ion milling for surface cleaning. (a) AFM micrograph of a Josephson junction after stripping the resist in acetone. (b) The same junction as in (a) after ion milling the surface with neon (p$_{\mathrm{Ne}} = 1\times10^{-4}$~mbar, E$_{\mathrm{Ne}} = 750$~eV) for 20 minutes in the UHV system. (c), (d) Simulated sputter rate for an ion current of 1~nA/mm$^2$ as a function of angle (c) and ion energy (d).  Average depth reached by the neon ions is shown as dashed lines. }
\end{figure} 
 \emph{iii) Shallow-angle ion milling} of surfaces with neutral atoms (e.g. argon or neon) is used in fabrication processes to remove surface contamination layers such as resist residues or surface oxides. As an example,  fabrication of superconducting qubits frequently involves ion milling to remove niobium or aluminium oxides before shadow evaporation of the first aluminum layer of the Josephson junction. This ion milling step was in the past found to reduce Q-values of qubits and resonators which led to the development of more elaborate alternative fabrication routes~\cite{17DunsworthAA, 19NersisyanAA}. The ion gun in the UHV treatment system allows to study the impact of ion bombardment with various ionic species and parameters. A typical process pressure of $1\times10^{-4}$~mbar is used during ion milling. For neon this can be done in extremely clean static-vacuum conditions. In contrast to argon and most contaminant species, neon is not pumped by the cryogenic titanium sublimation pump and only needs to be admitted into the chamber at the beginning of the process. The sputter system can also be used to investigate more reactive ionic species present in semiconductor etching procedures such as those from Cl$_2$, HBr, SF$_6$ or  CF$_4$ based processes. Last, we note that the ion beam can be adjusted in size and position on the sample. This could enable local treatments of the chip surfaces e.g. for selective trimming of quantum devices if a proper alignment system were to be implemented. 

To demonstrate the effectiveness of ion milling Figure~\ref{fig:fig5} shows two atomic force microscope images with significant resist residues before (a) and a clean surface  (b) after ion milling with neon at 0.75~kV for 20~min. The exact ion milling rate depends on flux, energy, material composition and beam angle and is hard to determine accurately in experiment. Instead we plot simulation results for neon that were obtained with the SRIM package~\cite{10ZieglerAA} for the most common substrate materials that occur in conventional Josephson Junction devices. Fig.~\ref{fig:fig5}(c) shows the dependence of the sputter rate on incident angle assuming a fixed ion flux of 1~nA/mm$^2$. The effective sputtering yield increases as the angle deviates further from the surface normal and has a maximum at around 60~$^\circ$.  For larger angles the yield is suppressed to zero as the beam becomes parallel to the sample surface. At the same time ion penetration depth and associated defect formation is reduced at shallower ion milling angles. This led to the choice of a 60~$^\circ$ incident angle which we believe to be optimal. The sputter rate depends on the used ion species and also varies by over a factor three between the different substrate materials. At a neon ion energy of 1.5~keV we measure an ion flux of 7~nA/mm$^2$ and estimate a sputter rate of 10-25~nm/h for the different substrate materials. This is roughly compatible with a value of 12~nm/h estimated from atomic force microscope images before and after sputtering. It is, however, difficult to differentiate between the sputter rate of different materials in experiment. Fig.~\ref{fig:fig5}(d) gives the simulated sputter rate as a function of ion energy for a fixed incident angle of 60~$^\circ$ for future reference.

\section{Qubit Measurements}
\begin{figure}[ht!]
\includegraphics[width=1.\columnwidth]{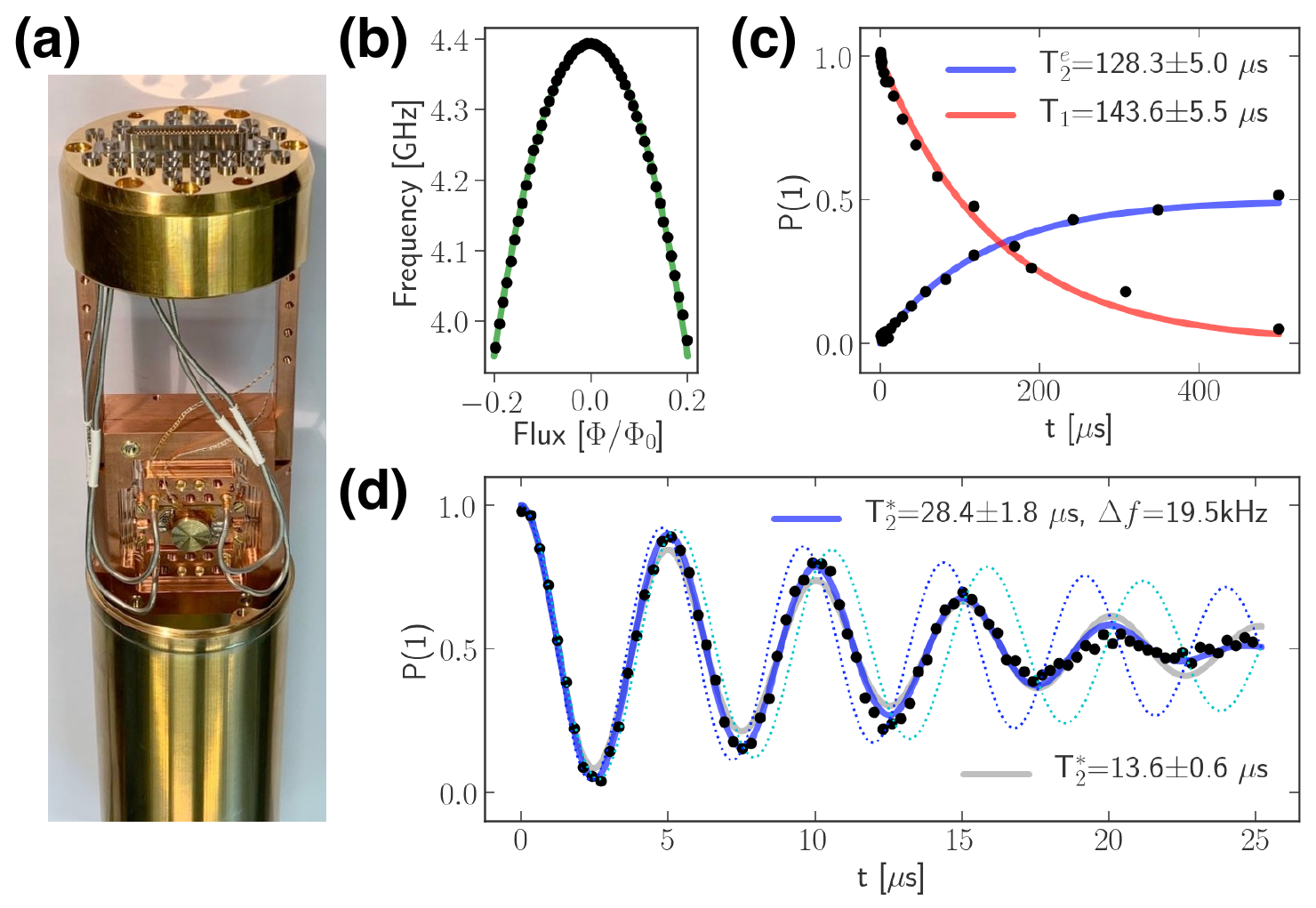}
\caption{\label{fig:fig6} (a) Image of the UHV package when loaded in the puck of a Bluefors bottom loading dilution refrigerator. (b) Qubit spectroscopy of a qubit chip inside the UHV package. (c) High-coherence qubit data with fits for $T_2^{echo}$ and $T_1$. (d) Ramsey measurement with single(two)-frequency fits in grey(blue) respectively giving correspondingly different $T_2^{*}$ values. Dashed lines indicate the two curves with a detuning $\Delta f$ that are averaged to obtain the blue curve.}
\end{figure}

The UHV-package was designed to be compatible with a fast-turnaround bottom loading dilution refrigerator (Bluefors LD). Fig.~\ref{fig:fig6}(a) shows a picture of the UHV-package fully connected and ready to be inserted into the probe of the corresponding loader mechanism. In this picture there is a single coil on the back of the UHV-package which is connected with four redundant wires to the 100-pin DC connector and from there with a superconducting loom to the top of the fridge. Four of the 36 microwave cables are attached to the sample inside the package.

As a real-world test of the UHV-package and the treatment system we loaded two chips with four flux tunable transmon qubits each. They are addressable through a common transmission line and $\lambda/2$ hanger resonators with slightly offset frequencies. The qubits are fabricated on an intrinsic high-resistivity Si substrate, where the resonators and qubit capacitors are made of niobium. The SQUID loops of the qubits were deposited via double angle evaporation of Al and in-situ oxidation. The chips are clamped in place between the copper pusher block and the PCB and then wire bonded using Al wire. The four microwave cables are connected to inputs and outputs of the transmission lines on each chip. Fig.~\ref{fig:fig6}(b) shows two-tone spectroscopy of a single qubit vs. applied external magnetic flux after loading the chips under UHV conditions. Here, the black dots are measured data points, whereas the lines represent analytical fits. Typical qubit frequencies are f$_{01}=4.4$~GHz at the sweetspot and anharmonicities $\alpha = 2($f$_{02}/2-$f$_{01})=-320$~MHz. Fig.~\ref{fig:fig6}(c) shows T$_1$ relaxation and Hahn echo measurements for one of the better qubits at the sweet spot. Fits to an exponential decay result in T$_1=143.6\pm5.5~\mu$s and T$^{echo}_2=128.3\pm5~\mu$s. The average over all eight qubits in the package was T$_1=84.3~\mu$s and T$^{echo}_2=134.4~\mu$s. Finally, Fig.~\ref{fig:fig6}(d) shows a typical Ramsey measurement. Instead of the expected simple damped oscillation $0.5+0.5~$cos$(\Delta\omega t)$e$^{-\mathrm{t/T}_2^*}$ we frequently find a beating in such measurements that seems to come from alternation between two dominant frequency detunings during a measurement. This may originate either in the presence of two-level fluctuators~\cite{17MullerAB} or quasiparticles that lead to parity fluctuations~\cite{12SunAA}. In  Fig.~\ref{fig:fig6}(d) we fit the measured curves both with a single frequency fit (grey) and a two-frequency fit (blue). The latter results in a $2\times$ longer T$_2^*$ and an extracted $\Delta f=19$~kHz.  Given an E$_J$/E$_C$-ratio of 27.3 and a corresponding residual charge dispersion of $\epsilon_0=23$~kHz, our findings are compatible with the presence of quasiparticles but do not exclude two-level fluctuators as the source of the observed fluctuations. The measured coherence figures are on par with the best in literature and show that our UHV-package can be used to measure complex quantum device chips and study a variety of surface treatments~\cite{2020MME}.

\section{Conclusions}
We presented a UHV treatment system and matching UHV package that allow the surface of quantum chips to be cleaned and passivated before loading them into a dilution refrigerator. The UHV system allows for annealing, ultra-violet light irradiation, ion milling and surface passivation. The UHV package was shown to retain vacuum to better than a few times $1\times10^{-8}$~mbar during the transfer and cool down of the samples with the help of a titanium getter layer. Measurements on high-coherence flux tunable qubits (T$_1=84.3~\mu$s and T$^{echo}_2=134.4~\mu$s) demonstrate that the UHV package is capable of assessing coherence figures of merit as a function of treatment parameters for complex multi-qubit chips. This functionality is unique in the field and will be put to use in future work~\cite{2020MME}.

\acknowledgments{We thank J. Chow for insightful discussions and R. Heller, H. Steinauer, A. Zulji and S. Gamper for technical support.
A.F., P.M. and S.F. acknowledge support form IARPA LogiQ program under contract W911NF-16-1-0114-FE for design and characterization of the UHV package.
A.F. and C.M. acknowledge support from the Swiss National Science Foundation through NCCR QSIT.}

\section*{Data Availability}
The data that support the findings of this study are available from the corresponding author upon reasonable request.


\begin{thebibliography}{14}%
\makeatletter
\providecommand \@ifxundefined [1]{%
 \@ifx{#1\undefined}
}%
\providecommand \@ifnum [1]{%
 \ifnum #1\expandafter \@firstoftwo
 \else \expandafter \@secondoftwo
 \fi
}%
\providecommand \@ifx [1]{%
 \ifx #1\expandafter \@firstoftwo
 \else \expandafter \@secondoftwo
 \fi
}%
\providecommand \natexlab [1]{#1}%
\providecommand \enquote  [1]{``#1''}%
\providecommand \bibnamefont  [1]{#1}%
\providecommand \bibfnamefont [1]{#1}%
\providecommand \citenamefont [1]{#1}%
\providecommand \href@noop [0]{\@secondoftwo}%
\providecommand \href [0]{\begingroup \@sanitize@url \@href}%
\providecommand \@href[1]{\@@startlink{#1}\@@href}%
\providecommand \@@href[1]{\endgroup#1\@@endlink}%
\providecommand \@sanitize@url [0]{\catcode `\\12\catcode `\$12\catcode
  `\&12\catcode `\#12\catcode `\^12\catcode `\_12\catcode `\%12\relax}%
\providecommand \@@startlink[1]{}%
\providecommand \@@endlink[0]{}%
\providecommand \url  [0]{\begingroup\@sanitize@url \@url }%
\providecommand \@url [1]{\endgroup\@href {#1}{\urlprefix }}%
\providecommand \urlprefix  [0]{URL }%
\providecommand \Eprint [0]{\href }%
\providecommand \doibase [0]{http://dx.doi.org/}%
\providecommand \selectlanguage [0]{\@gobble}%
\providecommand \bibinfo  [0]{\@secondoftwo}%
\providecommand \bibfield  [0]{\@secondoftwo}%
\providecommand \translation [1]{[#1]}%
\providecommand \BibitemOpen [0]{}%
\providecommand \bibitemStop [0]{}%
\providecommand \bibitemNoStop [0]{.\EOS\space}%
\providecommand \EOS [0]{\spacefactor3000\relax}%
\providecommand \BibitemShut  [1]{\csname bibitem#1\endcsname}%
\let\auto@bib@innerbib\@empty
\bibitem [{\citenamefont {Oliver}\ and\ \citenamefont
  {Welander}(2013)}]{13OliverAA}%
  \BibitemOpen
  \bibfield  {author} {\bibinfo {author} {\bibfnamefont {W.~D.}\ \bibnamefont
  {Oliver}}\ and\ \bibinfo {author} {\bibfnamefont {P.~B.}\ \bibnamefont
  {Welander}},\ }\href@noop {} {\bibfield  {journal} {\bibinfo  {journal} {MRS
  Bulletin}\ }\textbf {\bibinfo {volume} {38}},\ \bibinfo {pages} {816}
  (\bibinfo {year} {2013})}\BibitemShut {NoStop}%
\bibitem [{\citenamefont {M{\"u}ller}, \citenamefont {Cole},\ and\
  \citenamefont {Lisenfeld}(2017)}]{17MullerAB}%
  \BibitemOpen
  \bibfield  {author} {\bibinfo {author} {\bibfnamefont {C.}~\bibnamefont
  {M{\"u}ller}}, \bibinfo {author} {\bibfnamefont {J.~H.}\ \bibnamefont
  {Cole}}, \ and\ \bibinfo {author} {\bibfnamefont {J.}~\bibnamefont
  {Lisenfeld}},\ }\href@noop {} {\bibfield  {journal} {\bibinfo  {journal}
  {Rep. Prog. Phys.}\ }\textbf {\bibinfo {volume} {82}},\ \bibinfo {pages}
  {124501} (\bibinfo {year} {2017})}\BibitemShut {NoStop}%
\bibitem [{\citenamefont {Martinis}\ and\ \citenamefont
  {Megrant}(2014)}]{14MartinisAB}%
  \BibitemOpen
  \bibfield  {author} {\bibinfo {author} {\bibfnamefont {J.~M.}\ \bibnamefont
  {Martinis}}\ and\ \bibinfo {author} {\bibfnamefont {A.}~\bibnamefont
  {Megrant}},\ }\href@noop {} {\bibfield  {journal} {\bibinfo  {journal} {Arxiv
  preprint arXiv: 1410.5793}\ } (\bibinfo {year} {2014})}\BibitemShut {NoStop}%
\bibitem [{\citenamefont {Paladino}\ \emph {et~al.}(2013)\citenamefont
  {Paladino}, \citenamefont {Galperin}, \citenamefont {Falci},\ and\
  \citenamefont {Altshuler}}]{13PaladinoAA}%
  \BibitemOpen
  \bibfield  {author} {\bibinfo {author} {\bibfnamefont {E.}~\bibnamefont
  {Paladino}}, \bibinfo {author} {\bibfnamefont {Y.~M.}\ \bibnamefont
  {Galperin}}, \bibinfo {author} {\bibfnamefont {G.}~\bibnamefont {Falci}}, \
  and\ \bibinfo {author} {\bibfnamefont {B.~L.}\ \bibnamefont {Altshuler}},\
  }\href@noop {} {\bibfield  {journal} {\bibinfo  {journal} {Reviews of Modern
  Physics}\ }\textbf {\bibinfo {volume} {86}},\ \bibinfo {pages} {361}
  (\bibinfo {year} {2013})}\BibitemShut {NoStop}%
\bibitem [{\citenamefont {Ithier}\ \emph {et~al.}(2005)\citenamefont {Ithier},
  \citenamefont {Collin}, \citenamefont {Joyez}, \citenamefont {Meeson},
  \citenamefont {Vion}, \citenamefont {Esteve}, \citenamefont {Chiarello},
  \citenamefont {Shnirman}, \citenamefont {Makhlin}, \citenamefont {Schriefl},\
  and\ \citenamefont {et~al.}}]{05IthierAA}%
  \BibitemOpen
  \bibfield  {author} {\bibinfo {author} {\bibfnamefont {G.}~\bibnamefont
  {Ithier}}, \bibinfo {author} {\bibfnamefont {E.}~\bibnamefont {Collin}},
  \bibinfo {author} {\bibfnamefont {P.}~\bibnamefont {Joyez}}, \bibinfo
  {author} {\bibfnamefont {P.~J.}\ \bibnamefont {Meeson}}, \bibinfo {author}
  {\bibfnamefont {D.}~\bibnamefont {Vion}}, \bibinfo {author} {\bibfnamefont
  {D.}~\bibnamefont {Esteve}}, \bibinfo {author} {\bibfnamefont
  {F.}~\bibnamefont {Chiarello}}, \bibinfo {author} {\bibfnamefont
  {A.}~\bibnamefont {Shnirman}}, \bibinfo {author} {\bibfnamefont
  {Y.}~\bibnamefont {Makhlin}}, \bibinfo {author} {\bibfnamefont
  {J.}~\bibnamefont {Schriefl}}, \ and\ \bibinfo {author} {\bibnamefont
  {et~al.}},\ }\href@noop {} {\bibfield  {journal} {\bibinfo  {journal}
  {Physical Review B}\ }\textbf {\bibinfo {volume} {72}},\ \bibinfo {pages}
  {134519} (\bibinfo {year} {2005})}\BibitemShut {NoStop}%
\bibitem [{\citenamefont {de~Graaf}\ \emph {et~al.}(2018)\citenamefont
  {de~Graaf}, \citenamefont {Faoro}, \citenamefont {Burnett}, \citenamefont
  {Adamyan}, \citenamefont {Tzalenchuk}, \citenamefont {Kubatkin},
  \citenamefont {Lindstr\"om},\ and\ \citenamefont {Danilov}}]{18GraafAA}%
  \BibitemOpen
  \bibfield  {author} {\bibinfo {author} {\bibfnamefont {S.~E.}\ \bibnamefont
  {de~Graaf}}, \bibinfo {author} {\bibfnamefont {L.}~\bibnamefont {Faoro}},
  \bibinfo {author} {\bibfnamefont {J.}~\bibnamefont {Burnett}}, \bibinfo
  {author} {\bibfnamefont {A.~A.}\ \bibnamefont {Adamyan}}, \bibinfo {author}
  {\bibfnamefont {A.~Y.}\ \bibnamefont {Tzalenchuk}}, \bibinfo {author}
  {\bibfnamefont {S.~E.}\ \bibnamefont {Kubatkin}}, \bibinfo {author}
  {\bibfnamefont {T.}~\bibnamefont {Lindstr\"om}}, \ and\ \bibinfo {author}
  {\bibfnamefont {A.~V.}\ \bibnamefont {Danilov}},\ }\href@noop {} {\bibfield
  {journal} {\bibinfo  {journal} {Nature Communications}\ }\textbf {\bibinfo
  {volume} {8}},\ \bibinfo {pages} {1143} (\bibinfo {year} {2018})}\BibitemShut
  {NoStop}%
\bibitem [{\citenamefont {Koebley}, \citenamefont {Outlaw},\ and\ \citenamefont
  {Dellwo}(2012)}]{12KoebleyAA}%
  \BibitemOpen
  \bibfield  {author} {\bibinfo {author} {\bibfnamefont {S.~R.}\ \bibnamefont
  {Koebley}}, \bibinfo {author} {\bibfnamefont {R.~A.}\ \bibnamefont {Outlaw}},
  \ and\ \bibinfo {author} {\bibfnamefont {R.~R.}\ \bibnamefont {Dellwo}},\
  }\href@noop {} {\bibfield  {journal} {\bibinfo  {journal} {{Journal of Vacuum
  Science \& Technology A: Vacuum, Surfaces, and Films}}\ }\textbf {\bibinfo
  {volume} {30}},\ \bibinfo {pages} {060601} (\bibinfo {year}
  {2012})}\BibitemShut {NoStop}%
\bibitem [{\citenamefont {Foerster}, \citenamefont {Halama},\ and\
  \citenamefont {Lanni}(1990)}]{90FoersterAA}%
  \BibitemOpen
  \bibfield  {author} {\bibinfo {author} {\bibfnamefont {C.~L.}\ \bibnamefont
  {Foerster}}, \bibinfo {author} {\bibfnamefont {H.}~\bibnamefont {Halama}}, \
  and\ \bibinfo {author} {\bibfnamefont {C.}~\bibnamefont {Lanni}},\
  }\href@noop {} {\bibfield  {journal} {\bibinfo  {journal} {{Journal of Vacuum
  Science \& Technology A: Vacuum, Surfaces, and Films}}\ }\textbf {\bibinfo
  {volume} {8}},\ \bibinfo {pages} {2856} (\bibinfo {year} {1990})}\BibitemShut
  {NoStop}%
\bibitem [{\citenamefont {Kumar}\ \emph {et~al.}(2016)\citenamefont {Kumar},
  \citenamefont {Sendelbach}, \citenamefont {Beck}, \citenamefont {Freeland},
  \citenamefont {Wang}, \citenamefont {Wang}, \citenamefont {Yu}, \citenamefont
  {Wu}, \citenamefont {Pappas},\ and\ \citenamefont {McDermott}}]{16KumarAA}%
  \BibitemOpen
  \bibfield  {author} {\bibinfo {author} {\bibfnamefont {P.}~\bibnamefont
  {Kumar}}, \bibinfo {author} {\bibfnamefont {S.}~\bibnamefont {Sendelbach}},
  \bibinfo {author} {\bibfnamefont {M.~A.}\ \bibnamefont {Beck}}, \bibinfo
  {author} {\bibfnamefont {J.~W.}\ \bibnamefont {Freeland}}, \bibinfo {author}
  {\bibfnamefont {Z.}~\bibnamefont {Wang}}, \bibinfo {author} {\bibfnamefont
  {H.}~\bibnamefont {Wang}}, \bibinfo {author} {\bibfnamefont {C.~C.}\
  \bibnamefont {Yu}}, \bibinfo {author} {\bibfnamefont {R.~Q.}\ \bibnamefont
  {Wu}}, \bibinfo {author} {\bibfnamefont {D.~P.}\ \bibnamefont {Pappas}}, \
  and\ \bibinfo {author} {\bibfnamefont {R.}~\bibnamefont {McDermott}},\
  }\href@noop {} {\bibfield  {journal} {\bibinfo  {journal} {Phys. Rev.
  Applied}\ }\textbf {\bibinfo {volume} {6}},\ \bibinfo {pages} {041001}
  (\bibinfo {year} {2016})}\BibitemShut {NoStop}%
\bibitem [{\citenamefont {Dunsworth}\ \emph {et~al.}(2017)\citenamefont
  {Dunsworth}, \citenamefont {Megrant}, \citenamefont {Quintana}, \citenamefont
  {Chen}, \citenamefont {Barends}, \citenamefont {Burkett}, \citenamefont
  {Foxen}, \citenamefont {Chen}, \citenamefont {Chiaro}, \citenamefont
  {Fowler}, \citenamefont {Graff}, \citenamefont {Jeffrey}, \citenamefont
  {Kelly}, \citenamefont {Lucero}, \citenamefont {Mutus}, \citenamefont
  {Neeley}, \citenamefont {Neil}, \citenamefont {Roushan}, \citenamefont
  {Sank}, \citenamefont {Vainsencher}, \citenamefont {Wenner}, \citenamefont
  {White},\ and\ \citenamefont {Martinis}}]{17DunsworthAA}%
  \BibitemOpen
  \bibfield  {author} {\bibinfo {author} {\bibfnamefont {A.}~\bibnamefont
  {Dunsworth}}, \bibinfo {author} {\bibfnamefont {A.}~\bibnamefont {Megrant}},
  \bibinfo {author} {\bibfnamefont {C.}~\bibnamefont {Quintana}}, \bibinfo
  {author} {\bibfnamefont {Z.}~\bibnamefont {Chen}}, \bibinfo {author}
  {\bibfnamefont {R.}~\bibnamefont {Barends}}, \bibinfo {author} {\bibfnamefont
  {B.}~\bibnamefont {Burkett}}, \bibinfo {author} {\bibfnamefont
  {B.}~\bibnamefont {Foxen}}, \bibinfo {author} {\bibfnamefont
  {Y.}~\bibnamefont {Chen}}, \bibinfo {author} {\bibfnamefont {B.}~\bibnamefont
  {Chiaro}}, \bibinfo {author} {\bibfnamefont {A.}~\bibnamefont {Fowler}},
  \bibinfo {author} {\bibfnamefont {R.}~\bibnamefont {Graff}}, \bibinfo
  {author} {\bibfnamefont {E.}~\bibnamefont {Jeffrey}}, \bibinfo {author}
  {\bibfnamefont {J.}~\bibnamefont {Kelly}}, \bibinfo {author} {\bibfnamefont
  {E.}~\bibnamefont {Lucero}}, \bibinfo {author} {\bibfnamefont {J.~Y.}\
  \bibnamefont {Mutus}}, \bibinfo {author} {\bibfnamefont {M.}~\bibnamefont
  {Neeley}}, \bibinfo {author} {\bibfnamefont {C.}~\bibnamefont {Neil}},
  \bibinfo {author} {\bibfnamefont {P.}~\bibnamefont {Roushan}}, \bibinfo
  {author} {\bibfnamefont {D.}~\bibnamefont {Sank}}, \bibinfo {author}
  {\bibfnamefont {A.}~\bibnamefont {Vainsencher}}, \bibinfo {author}
  {\bibfnamefont {J.}~\bibnamefont {Wenner}}, \bibinfo {author} {\bibfnamefont
  {T.~C.}\ \bibnamefont {White}}, \ and\ \bibinfo {author} {\bibfnamefont
  {J.~M.}\ \bibnamefont {Martinis}},\ }\href@noop {} {\bibfield  {journal}
  {\bibinfo  {journal} {Appl. Phys. Lett.}\ }\textbf {\bibinfo {volume}
  {111}},\ \bibinfo {pages} {022601} (\bibinfo {year} {2017})}\BibitemShut
  {NoStop}%
\bibitem [{\citenamefont {Nersisyan}\ \emph {et~al.}(2019)\citenamefont
  {Nersisyan}, \citenamefont {Sete}, \citenamefont {Stanwyck}, \citenamefont
  {Bestwick}, \citenamefont {Reagor}, \citenamefont {Poletto}, \citenamefont
  {Alidoust}, \citenamefont {Manenti}, \citenamefont {Renzas}, \citenamefont
  {Bui}, \citenamefont {Vu}, \citenamefont {Whyland},\ and\ \citenamefont
  {Mohan}}]{19NersisyanAA}%
  \BibitemOpen
  \bibfield  {author} {\bibinfo {author} {\bibfnamefont {A.}~\bibnamefont
  {Nersisyan}}, \bibinfo {author} {\bibfnamefont {E.}~\bibnamefont {Sete}},
  \bibinfo {author} {\bibfnamefont {S.}~\bibnamefont {Stanwyck}}, \bibinfo
  {author} {\bibfnamefont {A.}~\bibnamefont {Bestwick}}, \bibinfo {author}
  {\bibfnamefont {M.}~\bibnamefont {Reagor}}, \bibinfo {author} {\bibfnamefont
  {S.}~\bibnamefont {Poletto}}, \bibinfo {author} {\bibfnamefont
  {N.}~\bibnamefont {Alidoust}}, \bibinfo {author} {\bibfnamefont
  {R.}~\bibnamefont {Manenti}}, \bibinfo {author} {\bibfnamefont
  {R.}~\bibnamefont {Renzas}}, \bibinfo {author} {\bibfnamefont {C.-V.}\
  \bibnamefont {Bui}}, \bibinfo {author} {\bibfnamefont {K.}~\bibnamefont
  {Vu}}, \bibinfo {author} {\bibfnamefont {T.}~\bibnamefont {Whyland}}, \ and\
  \bibinfo {author} {\bibfnamefont {Y.}~\bibnamefont {Mohan}},\ }in\ \href
  {\doibase 10.1109/IEDM19573.2019.8993458} {\emph {\bibinfo {booktitle} {2019
  IEEE International Electron Devices Meeting (IEDM)}}}\ (\bibinfo {year}
  {2019})\ pp.\ \bibinfo {pages} {31.1.1--31.1.4}\BibitemShut {NoStop}%
\bibitem [{\citenamefont {Ziegler}, \citenamefont {Ziegler},\ and\
  \citenamefont {Biersack}(2010)}]{10ZieglerAA}%
  \BibitemOpen
  \bibfield  {author} {\bibinfo {author} {\bibfnamefont {J.~F.}\ \bibnamefont
  {Ziegler}}, \bibinfo {author} {\bibfnamefont {M.~D.}\ \bibnamefont
  {Ziegler}}, \ and\ \bibinfo {author} {\bibfnamefont {J.~P.}\ \bibnamefont
  {Biersack}},\ }\href@noop {} {\bibfield  {journal} {\bibinfo  {journal}
  {Nuclear Instruments and Methods in Physics Research Section B}\ }\textbf
  {\bibinfo {volume} {268}},\ \bibinfo {pages} {1818} (\bibinfo {year}
  {2010})}\BibitemShut {NoStop}%
\bibitem [{\citenamefont {Sun}\ \emph {et~al.}(2012)\citenamefont {Sun},
  \citenamefont {DiCarlo}, \citenamefont {Reed}, \citenamefont {Catelani},
  \citenamefont {Bishop}, \citenamefont {Schuster}, \citenamefont {Johnson},
  \citenamefont {Yang}, \citenamefont {Frunzio}, \citenamefont {Glazman},
  \citenamefont {Devoret},\ and\ \citenamefont {Schoelkopf}}]{12SunAA}%
  \BibitemOpen
  \bibfield  {author} {\bibinfo {author} {\bibfnamefont {L.}~\bibnamefont
  {Sun}}, \bibinfo {author} {\bibfnamefont {L.}~\bibnamefont {DiCarlo}},
  \bibinfo {author} {\bibfnamefont {M.~D.}\ \bibnamefont {Reed}}, \bibinfo
  {author} {\bibfnamefont {G.}~\bibnamefont {Catelani}}, \bibinfo {author}
  {\bibfnamefont {L.~S.}\ \bibnamefont {Bishop}}, \bibinfo {author}
  {\bibfnamefont {D.~I.}\ \bibnamefont {Schuster}}, \bibinfo {author}
  {\bibfnamefont {B.~R.}\ \bibnamefont {Johnson}}, \bibinfo {author}
  {\bibfnamefont {G.~A.}\ \bibnamefont {Yang}}, \bibinfo {author}
  {\bibfnamefont {L.}~\bibnamefont {Frunzio}}, \bibinfo {author} {\bibfnamefont
  {L.}~\bibnamefont {Glazman}}, \bibinfo {author} {\bibfnamefont {M.~H.}\
  \bibnamefont {Devoret}}, \ and\ \bibinfo {author} {\bibfnamefont {R.~J.}\
  \bibnamefont {Schoelkopf}},\ }\href@noop {} {\bibfield  {journal} {\bibinfo
  {journal} {Phys Rev Lett}\ }\textbf {\bibinfo {volume} {108}},\ \bibinfo
  {pages} {230509} (\bibinfo {year} {2012})}\BibitemShut {NoStop}%
\bibitem [{\citenamefont {Mergenthaler}\ \emph {et~al.}(2020)\citenamefont
  {Mergenthaler}, \citenamefont {M\"uller}, \citenamefont {Ganzhorn},
  \citenamefont {Paredes}, \citenamefont {M\"uller}, \citenamefont {Salis},
  \citenamefont {Adiga}, \citenamefont {Brink}, \citenamefont {Sandberg},
  \citenamefont {Hertzberg}, \citenamefont {Filipp},\ and\ \citenamefont
  {Fuhrer}}]{2020MME}%
  \BibitemOpen
  \bibfield  {author} {\bibinfo {author} {\bibfnamefont {M.}~\bibnamefont
  {Mergenthaler}}, \bibinfo {author} {\bibfnamefont {C.}~\bibnamefont
  {M\"uller}}, \bibinfo {author} {\bibfnamefont {M.}~\bibnamefont {Ganzhorn}},
  \bibinfo {author} {\bibfnamefont {S.}~\bibnamefont {Paredes}}, \bibinfo
  {author} {\bibfnamefont {P.}~\bibnamefont {M\"uller}}, \bibinfo {author}
  {\bibfnamefont {G.}~\bibnamefont {Salis}}, \bibinfo {author} {\bibfnamefont
  {V.}~\bibnamefont {Adiga}}, \bibinfo {author} {\bibfnamefont
  {M.}~\bibnamefont {Brink}}, \bibinfo {author} {\bibfnamefont
  {M.}~\bibnamefont {Sandberg}}, \bibinfo {author} {\bibfnamefont
  {J.}~\bibnamefont {Hertzberg}}, \bibinfo {author} {\bibfnamefont
  {S.}~\bibnamefont {Filipp}}, \ and\ \bibinfo {author} {\bibfnamefont
  {A.}~\bibnamefont {Fuhrer}},\ }\href@noop {} {\bibfield  {journal} {\bibinfo
  {journal} {In Preparation}\ } (\bibinfo {year} {2020})}\BibitemShut {NoStop}%
\end{thebibliography}
\end{document}